\documentclass[letterpaper,pre,preprint,showpacs,12pt]{revtex4-1}
\usepackage{graphicx}
\graphicspath{{./}}

\newcommand{\boxit}[1]{\vbox{\hrule\hbox{\vrule\kern3pt\vbox{\kern3pt#1
                       \kern3pt}\kern3pt\vrule}\hrule}}

\newcommand{\plusorminus}{\pm}

\newcommand{\ie}{\textit{i.e.,}}
\newcommand{\eg}{\textit{eg.} }

\newcommand{\half}{\frac{1}{2}}

\newcommand{\notequal}{\neq}

\newcommand{\integral}{\int}

\newcommand{\infinity}{\infty}
\newcommand{\definedas}{\equiv}

\newcommand{\goesas}{\sim}

\newcommand{\lessthanorabout}
           {\mathrel{\raise.3ex\hbox{$<$\kern-.75em\lower1ex\hbox{$\sim$}}}}
\newcommand{\greaterthanorabout}
           {\mathrel{\raise.3ex\hbox{$>$\kern-.75em\lower1ex\hbox{$\sim$}}}}

\renewcommand{\vector}[1]{\vec #1}
\newcommand\tw[1]{{\color{red}[\kern -2pt[\lower 1pt\hbox{$_{\scriptscriptstyle TW}$} #1 ]\kern -2pt]}}
\newcommand\jq[1]{{\color{blue}[\kern -2pt[\lower 1pt\hbox{$_{\scriptscriptstyle JQ}$} #1 ]\kern -2pt]}}

\newcommand{\bbI}{{\bf 1}}
\newcommand{\bbC} {{\bf C}} 
\newcommand{\bbL}{{\bf L}}
\newcommand{\bbH}{{\bf H}}
\newcommand{\vV}{{\vector V}} 
\newcommand{\vQ} {{\vector Q}}
\newcommand{\vL} {{\vector \Lambda}}
 
\newcommand{\calE}{{\cal E}} 
\newcommand{\Gv}{\mathsf{G}}
\newcommand{\Pv}{\mathsf{P}}
\newcommand{\rv}{{\underline r}}
\newcommand{\vk}{\underline{\vector k}}
\newcommand{\transpose}{^{\rm T}}
\newcommand{\abs}[1]{\left |#1 \right |}
\newcommand{\etal}{{\em et al}}
\long\def\omitt#1{}

\begin{document}
\title{Singular electrostatic energy of nanoparticle clusters}
\author{Jian Qin}
\email{jianq@stanford.edu}
\affiliation{Institute for Molecular Engineering,
University of Chicago, Chicago, Illinois 60637}
\affiliation{Department of Chemical Engineering,
Stanford University, Stanford, California 94305}

\author{Nathan W. Krapf}
\affiliation{Department of Physics and James Franck Institute,
University of Chicago, Chicago, Illinois 60637}

\author{Thomas A.\ Witten}
\email{t-witten@uchicago.edu}
\affiliation{Department of Physics and James Franck Institute,
University of Chicago, Chicago, Illinois 60637}

\date{\today}
 
\begin{abstract}
The binding of clusters of metal nanoparticles is partly electrostatic.  We address difficulties in calculating the electrostatic energy when high charging energies limit the total charge to a single quantum, entailing unequal potentials on the particles.  We show that the energy at small separation $h$ has a singular logarithmic dependence on $h$.  We derive a general form for this energy in terms of the singular capacitance of two spheres in near contact $c(h)$, together with nonsingular geometric features of the cluster.   Using this form, we determine the energies of various clusters, finding that more compact clusters are more stable.  These energies are proposed to be significant for metal-semiconductor binary nanoparticle lattices found experimentally.   We sketch how these effects should dictate the relative abundances of metal nanoparticle clusters in nonpolar solvents.  
\end{abstract}
 \maketitle
\section{Introduction} \label{sec:introduction}

\begin{figure}[tbh]
\includegraphics[width=.9\hsize]{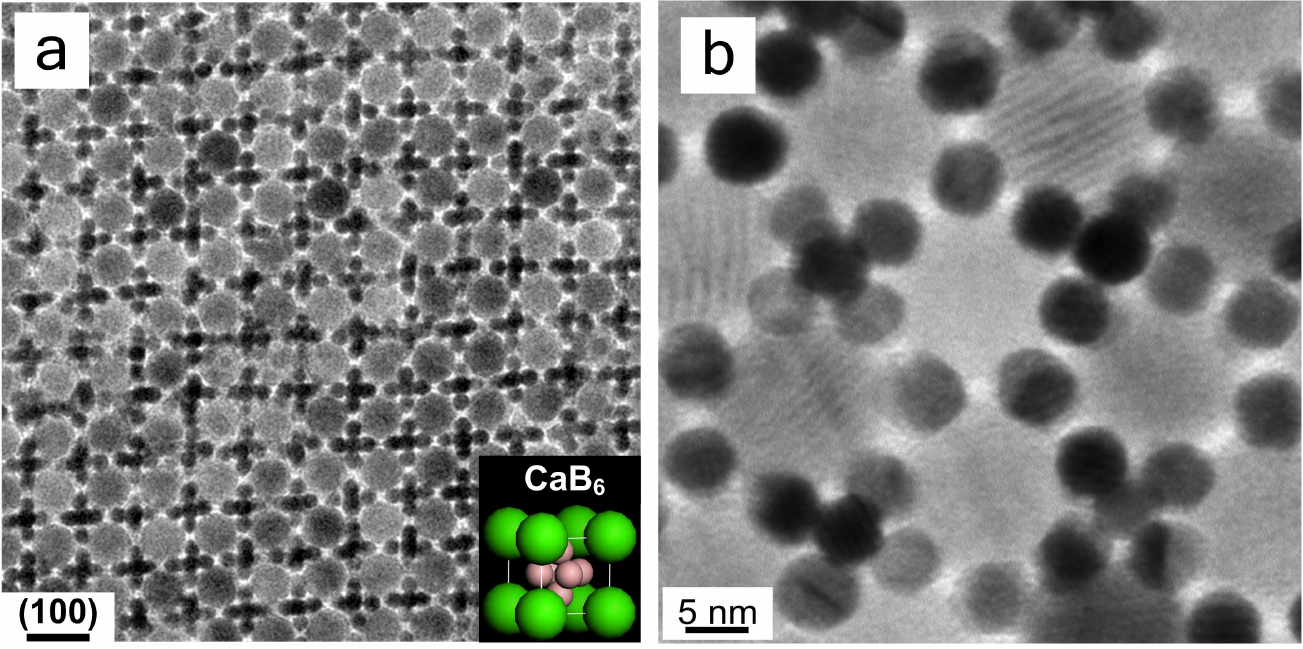}
\caption[]{ a: Transmission electron micrograph of experimental superlattice structure containing lead sulfate and dark colored palladium nanoparticles showing formation of regular palladium clusters as sketched in the colored inset.  Scale bar is 20 nm.  Reprinted by permission from Macmillan Publishers Ltd: from Ref.~\citenum{TalapinNature2006}, Fig.~1j, courtesy D. V. Talapin. \hfill\break
b: Transmission electron micrograph of a dodecagonal quasicrystal superlattice self-assembled from Fe$_2$ O$_3$ nanocrystals and clustered dark-colored 5-nm gold nanocrystals.  Reprinted by permission from Macmillan Publishers Ltd from Ref.~\citenum{Talapin:2009uq} Fig.~2b, courtesy D. V. Talapin.}
\label{fig:micrographs}
\end{figure}  

In self-assembled lattices of nanoparticles one often encounters clusters of metal particles \cite{Chen:2010fk}  as shown in Fig.~\ref{fig:micrographs}.  The remarkable stability of these clusters was argued to depend partly on states of nonzero electric charge \cite{TalapinNature2006}.  For particles of nanometer scale, such states are dominated by the quantization of charge.  The energy to add a single electron to a particle becomes large on the scale of the thermal energy $k_B T$, so that net charge on a particle is atypical.  Thus any net charge on a cluster is necessarily unevenly distributed over its particles. Still, a net charge on one particle must polarize the surrounding particles, producing electrostatic attraction.  This contrasts with the macroscopic case in which the available charge would be shared amongst the particles, producing repulsion.  It is of great interest to understand what types of clusters are favored under this simple and novel binding mechanism. Mutual electrostatic interactions between spherical conductors and with surfaces are of interest in space environments \cite{Boyer:1994kx} and in scanning probe microscopy \cite{Kalinin:2004rw}. Merrill \etal. \cite{Merrill:2009fk} explored the interactions among charged colloidal particles in clusters in solution. 

\begin{figure}
\hbox to \hsize {\hfill 
\includegraphics[width=.4\hsize]{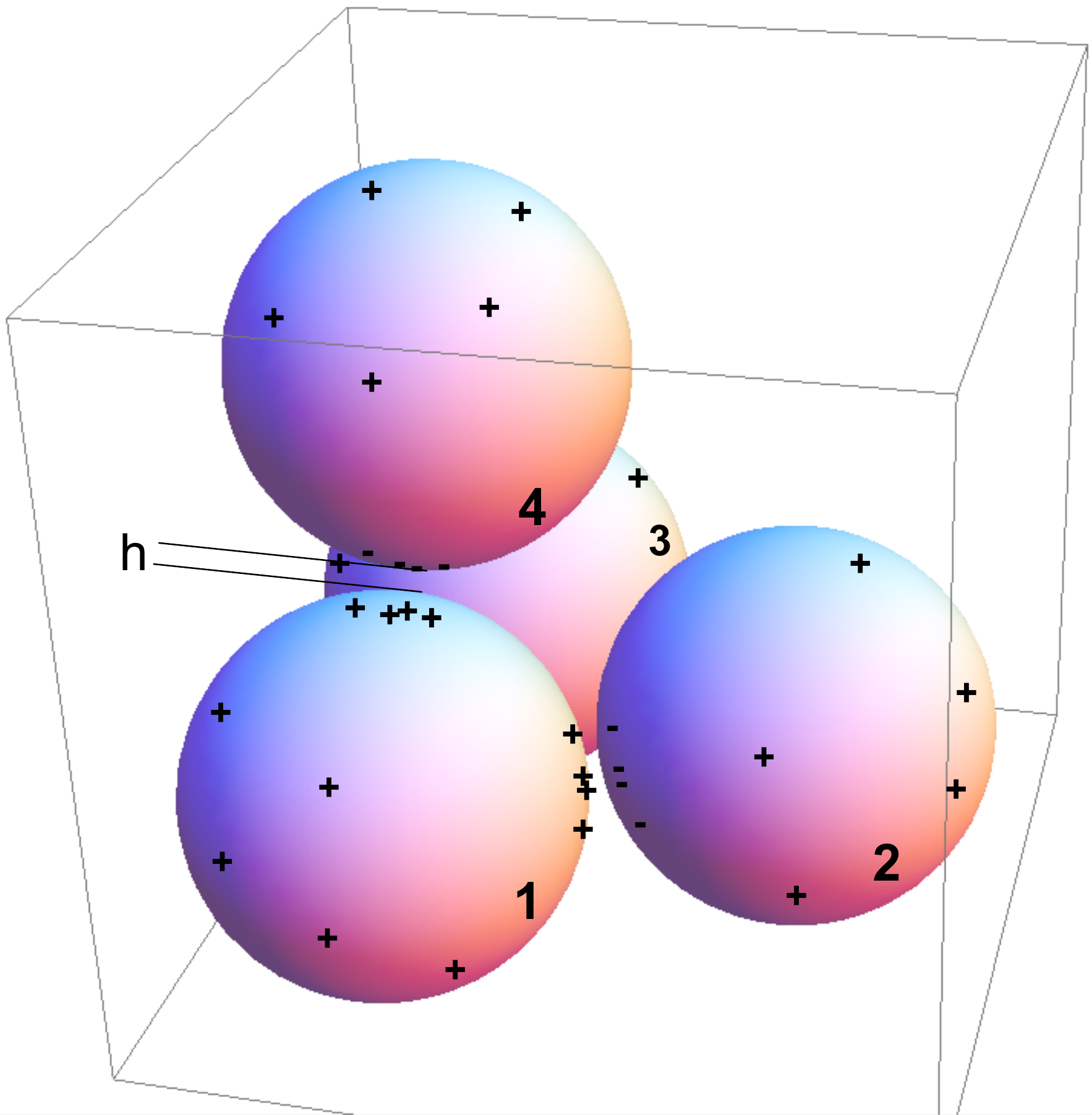}\hfill
\includegraphics[width=.4\hsize]{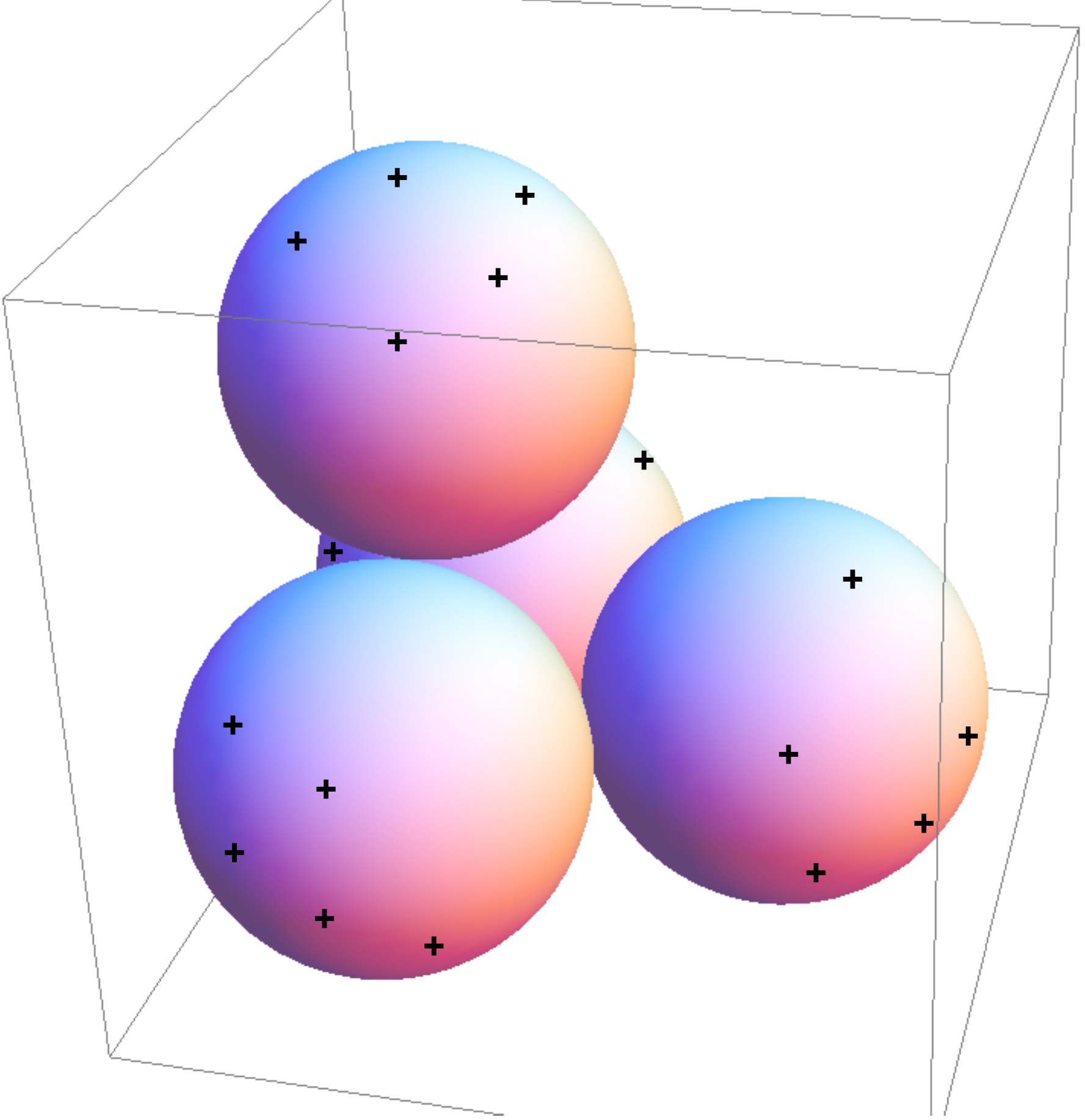}\hfill
}
\caption{Sketch of a cluster whose electrostatic energy is to be calculated. Sphere 1 has a net charge $Q$; spheres 2, 3, and 4 have no net charge.  Spheres 1, 2 and 3 are in near contact. Sphere 4 is in near contact with spheres 1 and 3 only.  Near contacts have separation $h$ much smaller than the sphere radii. Contact charges $q$ between spheres 1 and 2, and spheres 1 and 4 are shown.  Similar contact charges between spheres 1 and 3 and spheres 2 and 3 are hidden from view.  Right. Same cluster with the charge $Q$ free to migrate between spheres.  Spheres are at the same potential and there is no contact charge (cf. Fig.~\ref{fig:dimerEnergy}).  The regular tetrahedron treated in Fig.~\ref{fig:ETetra} is obtained by moving sphere 4 so that it contacts all the other spheres. }
\label{fig:sketches}
\end{figure}

Unlike most interactions of small particles, this electrostatic interaction cannot be reduced to a pairwise potential energy. Charge on one sphere induces polarization on each nearby sphere. This polarization induces further polarization in other spheres, as shown in Fig.~\ref{fig:sketches}. Since their separation is not large compared to their radius, the polarization cannot be accurately described by a dipole approximation.  Instead, all the spheres carry a polarization charge distribution that must be found self-consistently to minimize the electrostatic energy.  It is not known what types of clusters would be favored by this novel multi-body interaction mechanism. Moore \cite{AMooreArXiV2010} has provided a multipole formalism for calculating this energy and has explored the energies of simple clusters. Recently Qin and Freed \cite{Qin2015} provided a systematic method for determining electrostatic energies for polarizable insulating spheres using image charge methods.

\begin{figure}
\includegraphics[width=.8\textwidth]{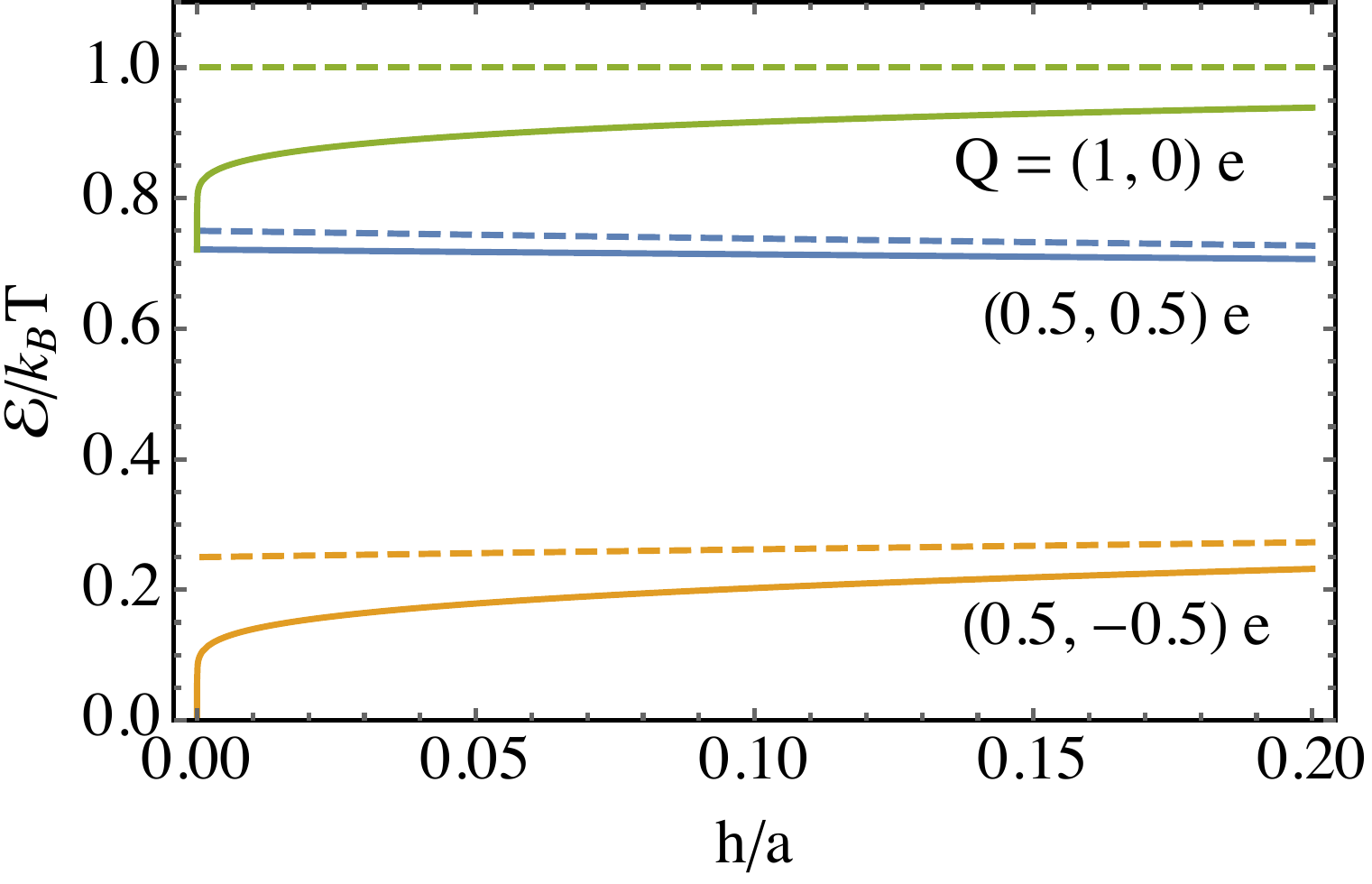}
\caption{\label{fig:energyScale} Coulomb energy $\calE$  vs normalized separation  for a pair of conducting spheres with radius $a = 27.8$ nm, bearing one quantum $e$ of total charge.  Radius was chosen to make the self energy of a single sphere equal to the thermal energy  $k_B T$ at room temperature.  Upper solid curve: all charge on one sphere.  Upper dashed curve: energy without charge polarization. Binding energy of about 0.24 $k_B T$ is largely due to the rapid decrease of $\calE$ at very small separation.  Middle curves: the charge is equally divided between the two spheres.  Lower curves: equal and opposite charges on the two spheres.  A change of sphere radius $a$ would change the energy scale in proportion to $1/a$.}
\label{fig:dimerEnergy}
\end{figure}

These polarization effects are nontrivial even for the case of two isolated spheres.  Numerical solutions by A. Russell \cite{Russell:1923xe}, by Pisler \etal. \cite{Pisler:1970zr} and by Kalinin \etal. \cite{Kalinin:2004rw} have been developed. A case of special interest is that of identical spheres of radius $a$ bearing equal and opposite charge $q$ at separation $h$. At small separation $h \ll a$ the charge becomes concentrated arbitrarily strongly near the contact point. This concentrated contact charge creates a logarithmically singular mutual capacitance  $c(h)$ of the form $c(h) \rightarrow \frac 1 4 a \log(\alpha a/h)$ (in electrostatic units \cite{Jackson}), where $\alpha$ is a numerical constant.  The resulting electrostatic energy $\half q^2/c(h)$ shown in Fig.~\ref{fig:dimerEnergy} reflects this singular behavior.  This divergent contact charge complicates the treatment of clusters of spheres with different charges. Moore's recent work on such clusters \cite{AMooreArXiV2010} shows a non-regular dependence of the energy on separation.  

Below we investigate the implications of the singular contact charge for the electrostatic energy $\calE(h)$ of clusters of conducting spheres $i$ at small separation $h$ when the total charge $Q$ resides on only one sphere, as sketched in Fig.~\ref{fig:sketches}.  
We contrast this energy with the simpler equipotential case where the charge $Q$ is allowed to pass freely between the spheres.  Then there is no contact charge, and the electrostatic energy $\calE_e(h)$ varies smoothly with $h$.   However in the case of interest where only one sphere is charged, new behavior arises owing to the appearance of contact charge. It is necessary to characterize this new behavior in order to find the desired electrostatic energy when the separations $h$ are small. We find that the energy at contact remains finite and equal to $\calE_e(0)$, but it acquires a logarithmic correction in $h$:
\begin{equation}
\calE(h)  \longrightarrow  \calE_e(0)~[1 + A/c(h) + \cdots ] ,
\label{eq:announce_result}
\end{equation} 
where the coefficient $A$ is independent of $h$ and depends only on position of the extra charge in the cluster,
on the equipotential charges and on the topological connectivity of the cluster.  

We begin by reviewing the origin of the singular $c(h)$ in Sec.~\ref{sec:twospheres}.
Next we define a capacitance matrix $\bbC(h)$ that gives the proportionality between the charges $Q_i$ and the potentials $V_i$ in Sec.~\ref{sec:cluster_energy}.
In Sec.~\ref{sec:asymmetrically}
we separate the regular and singular contributions to $\bbC(h)$ to obtain a parameterized expression for small $h$.
In Sec.~\ref{sec:energy}
we derive the form shown in Eq.~(\ref{eq:announce_result}), valid for asymptotically small $h$.
In Sec.~\ref{sec:geometry} we discuss how this energy is affected by extended versus compact cluster shapes.
Because the logarithmic singularity is weak,
the regular contributions to $\bbC(h)$ become significant for realistic contacts.
In Sec.~\ref{sec:examples},
we determine the leading regular contributions for several simple clusters using a straightforward numerical procedure.
Finally we comment on experimental implications and tests.
 
\section{Mutual capacitance of two spheres near contact}
\label{sec:twospheres}

For completeness we recall the origin of the logarithmic divergence of the mutual capacitance of two neighboring spheres of radius $a$, bearing equal and opposite charges $q$.  The potential difference between the spheres is denoted $V$.  In the limit $h/a \ll 1$, the capacitance is dominated by the adjacent sections of the two spheres.  Since the curvature there is very small on the scale of $h$, we may find the capacitance from this region via the Derjaguin approximation \cite{Safran:1994uq}.  This approximation treats the system as a set of concentric annular ring capacitors, neglecting the slopes of the surfaces within each ring.  At lateral distance $x$ from the central axis, where the separation is $y(x)$, the electric field $E$ is evidently $V/y(x)$.  Thus the surface charge density $\sigma(x) = E/(4\pi) = V/(4 \pi y(x))$.  To find the charge $q$, we integrate $\sigma$:
\begin{equation}
q = \integral 2\pi~x~dx~ \sigma(x).
\end{equation}
We note that the local height $y(x)$ is given by $x^2 + (a - (y - h)/2)^2 = a^2$ so that $2x~dx +  2 (a - (y - h)/2)~(-\half )dy = 0$.  Then for $a \gg y$,
\begin{equation}
2x~dx \rightarrow a~dy, 
\end{equation}
and
\begin{equation}
q \rightarrow \integral_h^{\alpha a}  2\pi V~ a~ dy / (4\pi y), 
\end{equation}
where $\alpha~ a$ is some upper cutoff of thickness where the Derjaguin approximation breaks down.
 Thus,
\begin{equation}
q \rightarrow \half V~a \integral_h^{\alpha a} dy/y = \half V~a~ \log(\alpha~ a/h)
\end{equation}
as claimed.  A change in the cutoff parameter $\alpha$ has no affect on the singular amplitude; it only adds a constant independent of $h$.  Thus any choice of $\alpha$ is equally valid for describing the singular behavior.  We shall arbitrarily take $\alpha=1$ below.    The capacitance  $q/(2V) \rightarrow \frac 1 4 a\log (a/h) + \hbox{const.}~$ thus goes logarithmically to infinity as $h\rightarrow 0$.  We define the singular part of this capacitance $\frac 1 4 a\log (a/h) \definedas c(h)$ for use below.

From this capacitance we can infer the energy needed to separate the contacting spheres with charges $\plusorminus q$.  At contact, the energy $\calE(0)$ is given by $\half q^2~(2V/q)$.  Since $(q/2V)\rightarrow \infinity$, we have vanishingly small $\calE$ at contact.  At infinite separation we have the full Coulomb self energy $2 ~\half q^2/a$.  Thus with equal and opposite charges the polarization of the spheres cancels virtually all the electrostatic energy of the separated spheres.
Fig.~\ref{fig:dimerEnergy} shows how this singularity influences the exact energy for this case.

\section{Capacitance matrix of a cluster}
\label{sec:cluster_energy}

We now extend our discussion of charges and potentials to a cluster of $n$ spheres labeled by $i$. We denote the set of charges $Q_i$ by the vector $\vQ$. There is in general a linear relationship between the charges $\vQ$ and the potentials on the spheres $\vV$ of the form 
\begin{equation}
\vQ = \bbC \vV, 
\end{equation} 
where $\bbC$ is  the $n\times n$ symmetric ``capacitance matrix".

In general the potentials $V_i$ are not equal, so that contacting spheres $i$ and $j$ acquire a singular contact charge on sphere $i$ at its contact with sphere $j$: $Q_{ij} = c(h) (V_i - V_j) $.  There is in general additional nonsingular charge $H_{ij} V_j$ for any sphere $j$ in the cluster, which remains finite as $h\rightarrow 0$.  The total charge on $i$ is then given by 
\begin{equation}
Q_i =\sum_j (c(h) (V_i - V_j)  + H_{ij} V_j)= c(h)\sum_{j(i)}  V_i - c(h) \sum_{j(i)} V_j + \sum_j H_{ij} V_j.
\end{equation}
Here the index $j(i)$ runs over all the spheres contacting sphere $i$.
The two terms in $c(h)$ can be expressed compactly in terms of the symmetric ``Laplace matrix" $\bbL$ defined by  $\bbL_{ij} = -1$ for all contacting spheres $i$ and $j$ and $\bbL_{ii} = -\sum_{j \notequal i} \bbL_{ij}$. Thus $\bbL_{ii}$ is the number of spheres contacting sphere $i$.  Likewise, we denote $\bbH$ as the matrix of $H_{ij}$'s. Thus
\begin{equation}
\vQ = (c(h) \bbL + \bbH ) \vV
\label{eq:bbLplusbbH}
\end{equation}
and so  $\bbC = c(h) \bbL + \bbH$.  
In this language we may readily express the electrostatic energy $\calE$ for any charge state $\vQ$: 
\begin{equation}
\calE(h) = \half \vQ \cdot \vV = \half \vQ \cdot {\bbC(h)}^{-1} \vQ .
\end{equation}

Since the $c(h)$ term depends only on potential differences, it vanishes whenever $\vV$ is uniform with potential $V_e$.  This ``equipotential state" is an important starting point for our derivation.  It is convenient to define a ``uniform vector'' $\vector u \definedas (1, 1, 1, \cdots 1)$.  Then in the equipotential state the potentials have the form $\vV \definedas  V_e ~\vector u$.  Since all spheres have the same potential, there are no contact charges, $\bbL \vector u = 0$ and $\bbC\vector u = \bbH \vector u$.  In general the charges $\vQ_e = V_e ~ \bbH \vector u$ for the equipotential cluster are not equal.   These charges $\vQ_e$ depend smoothly on $h$ with no singularity as $h\rightarrow 0$. The total charge $Q$ is given by $\vQ_e \cdot \vector u$.  For a given total charge $Q$, the potential $V_e$ is then given by $Q = \vector u \cdot \vQ = V_e ~\vector u \cdot \bbH \vector u$. Evidently the equipotential capacitance $C_e$ is simply $Q/V_e = \vector u \cdot \bbH \vector u$.  

It remains to determine how the singular $c(h)$ affects the $\vV$ and $\calE$ when the charges are different from $\vQ_e$.  We note that this problem bears a strong formal resemblance to that of determining contact forces in a weakly compressed mass of droplets \cite{Morse:1993fk}.

\section{Cluster with imposed charges}
\label{sec:asymmetrically}
When we specify the charges $\vQ \notequal \vQ_e$, the potentials must become unequal.  Then contact charges must appear, and the Laplacian matrix $\bbL$ becomes important.  
First we note that $\bbL \vV$ is nonzero for {\em all} nonuniform $\vV$ for the connected clusters considered here \cite{positiveNote}.
Thus in the limit of $c(h)\rightarrow \infinity$, Eq.~(\ref{eq:bbLplusbbH}) implies that any fixed, non-uniform $\vV$ creates diverging charges $\vQ$.  Only if $\vV$ becomes uniform can the charges be equal to the given charges.  That is, the potentials must approach the equipotential case treated above: $\vV \rightarrow V_e \vector u$, where $V_e$ is the equipotential voltage $Q/(\vector u \cdot \bbH \vector u)$.  For finite $c(h)$ we may separate $\vV$ into its limiting part $V_e \vector u$ plus a (small) remainder $\vV'$.  Likewise we may separate the charges $\vQ$ into the equipotential part $\vQ_e$ and a remainder $\vQ'$.  In this language Eq.~(\ref{eq:bbLplusbbH}) becomes
\begin{equation}
\vQ_e + \vQ' = \bbH (V_e \vector u + \vV') + c(h)~ \bbL (V_e \vector u + \vV').
\end{equation}  
Noting that $\bbL \vector u = 0$ and $V_e \bbH \vector u = \vQ_e$, this yields an implicit equation for the remainder potentials in terms of the known remainder charge:
\begin{equation}
\vQ' = \bbH \vV' + c(h)~ \bbL \vV' .
\label{eq:Q'V'}
\end{equation}
The uniform part of this equation can be found by forming the dot product with $\vector u$.  On the left side, $\vector u \cdot \vQ'$ vanishes by construction, since $\vQ_e$ contains the total charge.  On the right side the term in $\bbL$ vanishes giving
\begin{equation}
0 = \vector u \cdot \vQ' = \vector u \cdot \bbH \vV' + c(h)~ \vector u \cdot \bbL \vV' 
= \vQ_e \cdot \vV' ,
\label{eq:duality}
\end{equation} 
\ie\  $\vQ_e$ is orthogonal to $\vV'$.

As a mapping from the space of $\vV'$ to the space of $\vQ'$, $\bbL$ is invertible, since $\bbL \vV' \notequal 0$ for all $\vV'$.
To avoid confusion, we denote the $\bbL$ restricted to the $\vQ'$ and $\vV'$ space as $\tilde\bbL$. Further, 
Eq.~(\ref{eq:Q'V'}) is invertible for sufficiently large $c(h)$ \cite{invertibleNote}:
\begin{equation}
\vV' = (\bbH + c(h) \tilde\bbL)^{-1} \vQ' .
\label{eq:V'ofQ'}
\end{equation}
We recall (Sec. \ref{sec:twospheres}) that $c(h)$ was only defined up to an arbitrary additive constant.  Here we see that this arbitrariness has no physical impact.  If we add a constant $c_0$ to $c(h)$ and subtract $c_0 \tilde\bbL$ from $\bbH$, the equation is unchanged.  Thus for any choice of $c_0$ there is always a regular $\bbH$ for which Eq.~(\ref{eq:V'ofQ'}) is valid.

In terms of the small quantity $1/c(h)$, this may be written
\begin{equation}
\vV' = \frac 1{c(h)} \left( \bbI+\frac 1{c(h)} \tilde\bbL^{-1} \bbH \right )^{-1}\tilde\bbL^{-1} \vQ'  .
\label{eq:V'Q'2}
\end{equation}
We note that as $c \rightarrow \infinity$ for fixed $\vQ'$, the factor in $(\cdots)$ becomes unity, and the correction $\vV'$ becomes independent of $\bbH$.  Thus in this limit the only part of $\bbH$ that influences  the full $\vV$ is the equipotential part: $\bbH \vector u$.  

\section{Electrostatic energy}
\label{sec:energy}

Given this expression for the potential vector $\vV$, we may find the electrostatic energy $\calE$ for a given charge vector $\vQ$: $\calE = \half \sum_i Q_i V_i = \half \vQ \cdot \vV$.  In terms of the charge difference $\vQ' \definedas \vQ - \vQ_e$ and potentials $\vV'$, $\calE$ can be written using Eq.~(\ref{eq:V'Q'2})
\begin{equation}
\calE = \half (\vQ_e + \vQ')\cdot (V_e \vector u + \vV')
= \half Q V_e  + \half  V_e~\vQ' \cdot \vector u + \half \vQ_e \cdot \vV'  + \half \vQ'\cdot \vV' .
\label{eq:calEbig}
\end{equation}
The first term is simply the equipotential energy $\calE_e$.  As noted above, the second term vanishes because $\vQ_e$ was defined to have the total charge, leaving no net charge in $\vQ'$.   The third term was shown to vanish in Eq.~(\ref{eq:duality}).  Thus
\begin{equation}
\calE = \calE_e   + \half \vQ' \cdot \vV'
= \calE_e   +  \half ~\frac 1 {c(h)} \vQ' \cdot  \left ( \bbI + \frac 1 {c(h)}\tilde\bbL^{-1}\bbH\right )^{-1}\tilde\bbL^{-1} \vQ'  . 
\label{eq:calEfull}
\end{equation}

In the limit where $c(h)$ is so large that higher orders in $1/c(h)$ can be neglected,
this reduces to the form announced in Eq.~(\ref{eq:announce_result})
\begin{equation}
\calE(h) = \calE_e + \half ~\frac 1 {c(h)}~ \vQ' \cdot \tilde\bbL^{-1} \vQ' .
\label{eq:oldcalE}
\end{equation}
Once the equipotential charges are determined from $\bbH \vector u$, the entire dependence on the charge distribution is governed by the Laplacian matrix $\bbL$, with no further dependence on the geometry of the cluster.
In practice for $h$ values as large as a few percent this lowest-order expansion proves inaccurate for the examples studied in Sec. \ref{sec:examples}.  Thus the full matrix form of Eq.~(\ref{eq:calEfull}) is preferable.  This includes all the dependence on $c(h)$ but neglects the (presumed regular) dependence of $\bbH$ on $h$.  Since these are low-dimensional matrices, the needed operations are straightforward.

Eq.~(\ref{eq:calEfull}) shows that one may isolate the singular part of the electrostatic energy for a cluster of conducting spheres close to contact, using nonsingular quantities which can be readily computed numerically.  The energy at $h=0$ but without conductance between spheres is the same as for the equipotential case where conductance is allowed.  This means that the imposed distribution of charge among the spheres has no effect on the energy when  $h\rightarrow 0$.  Conversely, if $\vQ = \vQ_e$, then $\vQ' = 0$ and the energy obtained from  Eq.~(\ref{eq:calEfull}) is independent of $h$.  The actual change of $\calE$ with $h$ then arises only from the smooth and regular dependence of $\bbH$ on $h$, neglected in our treatment.  
Finally, if $\vQ \notequal \vQ_e$, but the total charge $Q=0$,
the leading $\calE_e$ part of the energy vanishes, and the entire energy goes to zero with $h$.
In leading order, the only aspect of the cluster that affects the energy is its connectivity.

The correction in $1/c(h)$ in Eq.~(\ref{eq:oldcalE}) is necessarily positive,
since both $\tilde\bbL$ and $\tilde\bbL^{-1}$ are positive definite \cite{positiveNote}.
As seen from Fig.~\ref{fig:energyScale} above,
this increase of energy can depend strongly on $h$. In general it depends on which sphere is charged. 

\section{Effect of cluster geometry}
\label{sec:geometry} 
A central question arising from this distinctive electrostatic effect is to understand how the shape of the cluster affects its binding energy $\calE(\infinity) - \calE(h)$.  Clusters with the largest $|\calE(h)|$ and the strongest binding are expected to be more abundant.  It is natural to ask whether $\calE$ favors compact clusters or extended ones.  In the limit $h\rightarrow 0$ there is a clear preference for extended clusters. Here the cluster is an equipotential and is energetically equivalent to a single conducting object with the given total charge.  The favored shape is thus that with highest capacitance to ground and largest spatial extent \cite{Jackson}.  A cluster of $n$ spheres thus has the strongest binding when extended out in a straight line.  

However, as shown above, any departure from the $h \rightarrow 0$ limit brings strong changes in $\calE$.  The derivative of $\calE$ with $h$ is infinite at $h=0$.  Thus even small nonzero $h$ can have a significant effect on the relative $\calE$ of different clusters.  Appendix \ref{sec:shape_appendix} argues that the binding penalty from the $1/c(h)$ correction in Eq.~(\ref{eq:oldcalE}) favors compact clusters over extended ones for large $n$.  Moreover, this correction can be strong enough that the net binding is stronger for more compact clusters.  

\section{Examples}
\label{sec:examples}
In this section we show explicitly how the two matrices $\bbL$ and $\bbH$ lead to the $h$-dependent Coulomb energy $\calE$ for several clusters of interest. We first consider a regular tetrahedron, planar 4-sphere clusters and a regular octahedron.  
\subsection{Determination of  $\bbL$}
The Laplacian matrix $\bbL$ for the tetrahedron is immediately apparent since each sphere is in contact with all three others.  Thus according to Sec.~\ref{sec:cluster_energy},
\begin{equation}
\bbL = \left [\matrix
{3 & -1 & -1 & -1 \cr
-1 & 3 & -1 & -1 \cr
-1 & -1 & 3 & -1 \cr
-1 & -1 & -1 & 3 \cr} \right ].
\end{equation}
For square, we may number the spheres in sequence around the perimeter.  Each sphere is then in contact with its predecessor and its successor, with no other contact:
\begin{equation}
\bbL = \left [\matrix
{2 & -1 & 0 & -1 \cr
-1 & 2 & -1 & 0 \cr
0 & -1 & 2 & -1 \cr
-1 & 0 & -1 & 2 \cr} \right ].
\end{equation}
If the square is collapsed into a rhombus, $\bbL$ remains unchanged until two of the opposite spheres---\eg 1 and 3---make contact to form a diamond shape.  Then 
\begin{equation}
\bbL = \left [\matrix
{3 & -1 & -1 & -1 \cr
-1 & 2 & -1 & 0 \cr
-1 & -1 & 3 & -1 \cr
-1 & 0 & -1 & 2 \cr} \right ].
\end{equation}
Likewise we can rotate sphere 2 out of the plane of the other three.  Again $\bbL$ remains unchanged until sphere 2 makes contact with Sphere 4. 

In an octahedron all six spheres are equivalent and each makes contact with four others, so that 
\begin{equation}
\bbL = \left [\matrix
{4 & -1 & -1 & -1 & -1 & 0\cr
-1 & 4 & -1 & -1 & 0 &  -1\cr
 -1 &-1 & 4  & 0 & -1 & -1\cr
-1 & -1& 0 & 4 & -1 & -1\cr
-1 & 0 & -1 & -1 & 4  & -1\cr
0 &-1 & -1 & -1 &-1 & 4  \cr} \right ]
\end{equation}

\subsection{Numerical determination of $\bbH$}
\label{sec:findbbH}
Eq.~(\ref{eq:calEfull}) requires us to determine the regular part of $\bbC$, namely $\bbH$.  We perform this determination by numerically computing the full capacitance matrix $\bbC$ for some small, nonzero $h$ and for the equipotential case. The equipotential case gives $\vQ_e$ and thus allows us to determine $\vQ'$ for a given $\vQ$.  Then we may find $\bbH$ using Eq.~(\ref{eq:bbLplusbbH}).  The $\bbH$ thus obtained depends on the separation chosen.  Our supposition that $\bbH$ is regular implies that this $\bbH(h)$ converges smoothly to an asymptotic value as $h\rightarrow 0$.  Our numerical results provide a test of this supposition.

Consider a cluster configuration with specified charge vector $\vQ$.
The actual surface charge distribution minimizes the total electrostatic energy.
To use this fact, we discretize sphere surfaces into $N$ small patches uniformly distributed over each sphere
and denote the patch charges by $\sigma_{i,\alpha}$.
Here $i = 1, 2, \cdots n$ labels spheres and $\alpha = 1, 2, \cdots N$ labels the patches.
The energy for given $\{\sigma_{i,\alpha}\}$ then reads
$\calE = \frac{1}{2} \sigma \cdot \Gv ~ \sigma =
\frac{1}{2} \sigma_{i,\alpha}~ G_{i,\alpha; j, \beta} ~\sigma_{j,\beta}$,
where 
\begin{equation}
G_{i, \alpha; j, \beta} = \abs{\rv_{i,\alpha} - \rv_{j,\beta}}^{-1}
\end{equation} 
is the Coulomb kernel between patches $(i,\alpha)$ and $(j,\beta)$, and $\rv_{i,\alpha}$ is the vector position of the patch ${i, \alpha}$. 
A diagonal entry of $\Gv$ evidently represents the Coulomb energy of a patch in isolation.   This energy depends on the size and shape of the patch.  The contribution of this self energy to the total energy becomes negligible when the number of patches $N$ becomes sufficiently large. In our calculation we have taken all patches to have a single self energy, adjusted to reproduce the known energy of an isolated sphere.   

To minimize the energy subject to the constraints of total charges on each sphere,
we introduce a projection matrix $\Pv$ of dimension $n \times nN$,
that maps charges from the space of patches to the space of spheres.
The entries $P_{ik}$ are non-vanishing and set to $1$ only if the $k$th patch belongs to the $i$th sphere.
Then the constraints on charge distributions are $\Pv ~ \sigma = \vQ$.
We note that $\Pv$ and its transpose $\Pv\transpose$ obey the relations
$\Pv~\Pv\transpose = N \bbI_{n}$ and
$\Pv\transpose~\Pv = \bbI_{nN}$,
where $\bbI_n$ and $\bbI_{nN}$ are identity matrices of dimensions $n$ and $nN$ respectively. 
\begin{figure}[bt]
\centering
\includegraphics[width=\textwidth,height=!]{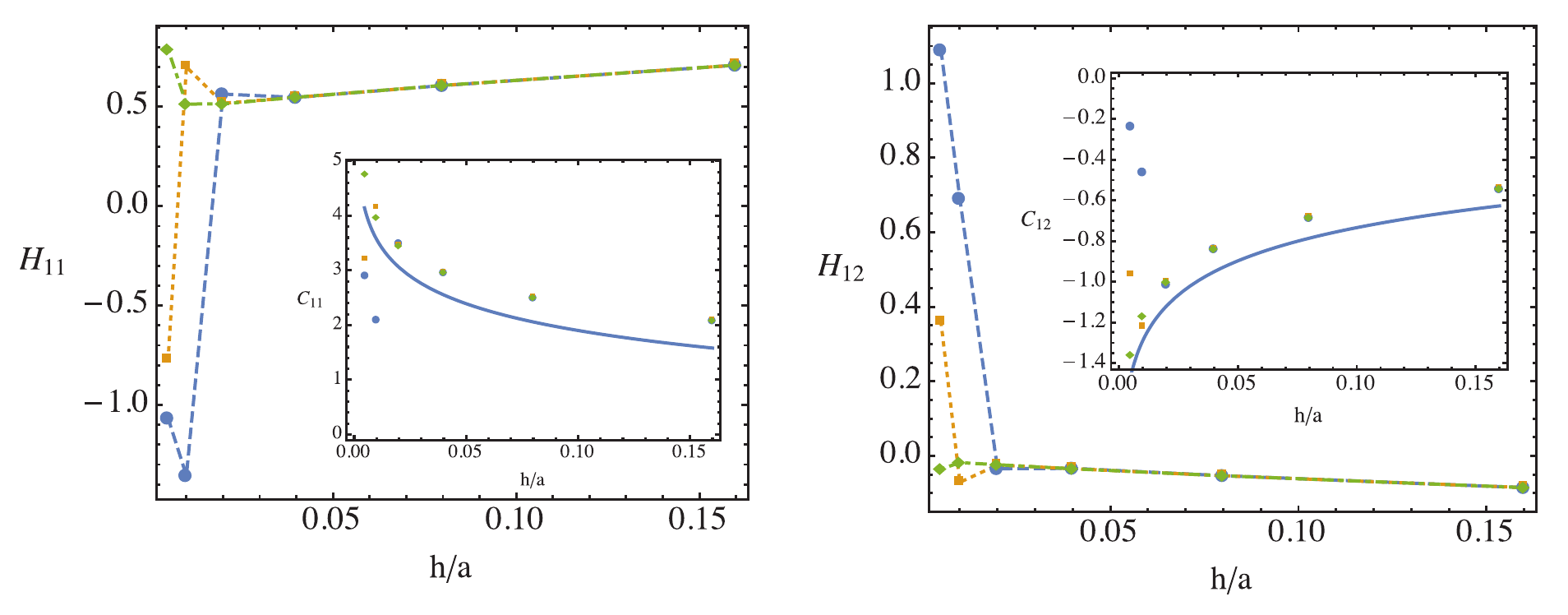}
\caption{Capacitance coefficients for a regular tetrahedron vs normalized separation $h/a$, determined as described in the text.  Left panel treats $C_{11}$; right panel treats $C_{12}$. The other eight elements of $\bbC$ are determined by symmetry.  Insets show the full capacitance, showing a strong dependence at small $h$ due to contact charge.  Main graphs show the regular parts $H_{11}$ and $H_{12}$ determined using Eq.~(\ref{eq:bbLplusbbH}).  Segmented lines connect the calculated values at different levels of discretization $N$: blue long dashes for $N=1000$, orange short dashes for $N=2000$ and green dot-dashes for $N=4000$. Irregular dependence for smallest $h$ is attributed to discretization errors and improves for finer discretization. The strong $h$ dependence of the inset has been removed and the different discretizations give a consistent extrapolation to $h=0$.  }
\label{fig:CTetra}
\end{figure}
\begin{table}[tbh]
\centering
\label{tab:tubescal}
\begin{tabular}{l | c   c   c   c   c   c | c}
\hline\hline
separation  &  $0$      &  $d$  &  $\sqrt{2}\,d$  &  $\sqrt{3}\,d$  & $1.6\,d$  & $1.9\,d$  &  $C_e$ \\
\hline
dimer           & $\ln2 + \gamma/2$&$-\gamma/2$  &  &          &          &          & $2\ln2$   \\
trimer-$\pi/2$  & \ $0.870$ &  $-0.193$ & $-0.249$  &          &          &          & $1.671$  \\
                &\ $1.036^a$&           &           &          &          &          &          \\
trimer-$\pi/3$  & \ $0.781$ &  $-0.121$ &           &          &          &          & $1.616$  \\
square          & \ $0.876$ &  $-0.158$ & $-0.101$  &          &          &          & $1.835$  \\
\hline
rhombus         & \ $0.783^b$ &  $-0.103^d$ &       & $-0.062$ &          &          & $1.823$  \\
                & \ $0.567^c$ & \ $0.036^e$ &       &          &          &          &          \\
\hline
tetrahedron     & \ $0.492$ &  $-0.017$ &           &          &          &          & $1.767$  \\
octahedron     & \ $0.228$ & \ $0.033$ & $-0.024$  &          &          &          & $2.024$  \\
cubic           & \ $0.703$ &  $-0.072$ & $-0.061$  & $-0.016$ &          &          & $2.308$ \\
icosahedron     &  $-0.019$ &  $-0.060$ &           &          & $-0.013$ & $-0.007$ & $2.509$ \\
dodecahedron    & \ $0.774$ &  $-0.091$ &           &          &          &          & $3.497$ \\
\hline\hline
\end{tabular}
\caption{
Elements of capacitance matrix $\bbH$ extrapolated to $h = 0$, tabulated by distance between spheres,
measured in terms of the sphere diameter $d$ in units of the capacitance of a single sphere.
Notes:  $a$: self energy of the sphere with one contact.
$b$: self energy of the sphere at the pointed site.
$c$: self energy of the sphere at the blunt site.
$d$: interaction between the pointed and blunt sites.
$e$: interaction between the two blunt sites.
For dodecahedron, the entries needed for distances $1.6 d$, $2.3d$, $2.6d$, and $2.8d$
are $-0.044$, $-0.007$, $-0.005$, and $-0.004$ respectively.
$\gamma = 0.577$ is the Euler gamma number. Last column shows the equipotential capacitance $C_e$ of the cluster.
}
\label{tab:Hregular}
\end{table}

We implement the constraint on $\vQ = \Pv \sigma$ by adding a Lagrange multiplier energy $\lambda_i \sum_\alpha \sigma_{i, \alpha}$ for each sphere $i$.  Defining $\vL = \{\lambda_1, \lambda_2, \cdots \lambda_n\}$, this amounts to minimizing $\half \sigma\cdot \Gv~ \sigma - \vL \cdot \vQ = \half \sigma\cdot \Gv~ \sigma - \vL \cdot \Pv \sigma$.
Setting the gradient $\partial / \partial \sigma_{i, \alpha}$ equal to zero yields the implicit equation for the  minimizing $\sigma$, denoted $\sigma^*$, in terms of $\vL$:
\begin{equation}
\Gv \sigma^* = \Pv\transpose \vL ~.
\label{eq:Gsigma}
\end{equation}
Now the minimizing energy $\calE^* = \half \sigma^* \cdot \Gv ~\sigma^*$ can be written as
\begin{equation}
\calE^* = \half \sigma^* \cdot \Pv\transpose \vL = \half (\Pv \sigma^*) \cdot \vL
= \half \vQ \cdot \vL ~.
\label{eq:calEmin}
\end{equation}
From this it is clear that $\vL$ is simply the set of potentials on the spheres $\vV$. 
We may obtain $\vQ$ in terms of $\vL$ using Eq.~(\ref{eq:Gsigma})
\begin{equation}
\vQ= \Pv \sigma^* = \Pv \Gv^{-1} \Pv\transpose \vL ~.
\end{equation}
Using $\vL = \vV$ and simplifying, 
\begin{equation}
\vQ= \Pv \sigma^* = (\Pv\Gv \Pv\transpose)^{-1} \vV.
\end{equation}
Evidently the capacitance matrix $\bbC$ is the matrix $(\Pv\Gv \Pv\transpose)^{-1}$.  Thus to determine $\bbC(h)$ it suffices to compute $\Gv$, project it to form the $n \times n$ matrix $\Pv\Gv \Pv\transpose$, and invert it.
$\bbH$ is then calculated by subtracting $c(h) \bbL$ from this $\bbC$.  

Figure~\ref{fig:CTetra} illustrates the results of this procedure for a regular tetrahedron.  Here $\bbC$ was computed for several small values of $h$.  The smallest $h$'s were comparable to the separation between the patches, so that discretization errors were significant.  Beyond this $h$ $\bbC$ showed the expected logarithmic singularity as in Fig. \ref{fig:energyScale}.  However, once $c(h) \bbL$ was subtracted to form $\bbH$, the $h$ dependence was gradual, smooth, and consistent for different discretizations.  Thus the expectation of smooth $\bbH$ was confirmed.  

We found similar confirmation for the $H_{ij}$ of other clusters.
The characteristic entries of $\bbH$ at $h=0$
for all clusters considered are tabulated in Table~\ref{tab:Hregular}.
Only the independent entries are shown.
The full capacitance matrix can be constructed by considering symmetry.
For all clusters considered, the agreements are excellent for $0 \leq h/a \leq 0.05$. 
\subsection{Cluster energies}
\label{sec:clusterEnergies}

\begin{figure}[bt]
\centering
\includegraphics[width=.5\textwidth,height=!]{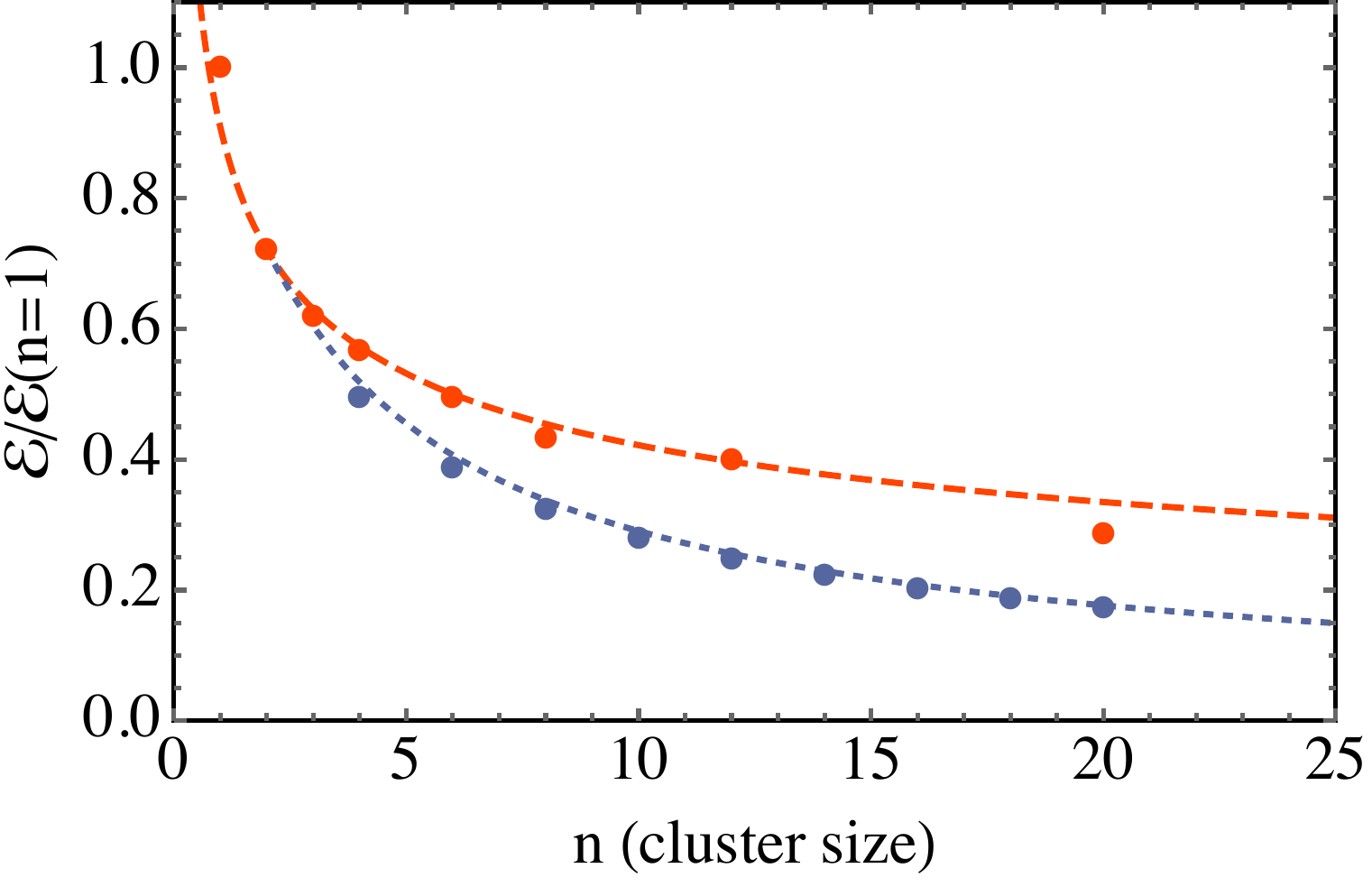}
\caption{
System size dependence of equipotential capacitance energy $\calE_e$ for a single sphere,
a dimer,  equilateral triangle,
a tetrahedron, an octahedron, a  cube, an icosahedron and a dodecahedron.
upper curve: $\calE = (2 / n)^{1/3} / ( 2 \ln(2) ) $. the energy for compact configurations.
lower curve: energy of cylinder of the same volume,
with length $L=2n\,a$ and radius $r_0 = 0.816\,a$:  $\calE =\ln(2n a/r_0)/(n \ln(2) \ln(2a/r_0) ) $ \cite{Maxwell1878}.
}
\label{fig:Esize}
\end{figure}

Table~\ref{tab:Hregular} also includes values for the equipotential capacitance $C_e = \vector u\cdot\bbH \vector u$, which gives the $h \rightarrow 0$ cluster binding energies,
$\calE_e = \frac{1}{2} Q^2/C_e$. This energy depends on the cluster size and shape but not at all on the charge placement.
As anticipated in Sec.~\ref{sec:geometry}, we found that $\calE_e$ increases with the number of spheres $n$,
as shown in Fig.~\ref{fig:Esize}.
For compact clusters, the data scale with the system size as $n^{-1/3}$ while for extended clusters $\calE_e \goesas \ln n/n$.  This expected behavior is discussed in Appendix \ref{sec:shape_appendix}.

\begin{figure}[bt]
\centering
\includegraphics[width=.45\textwidth,height=!]{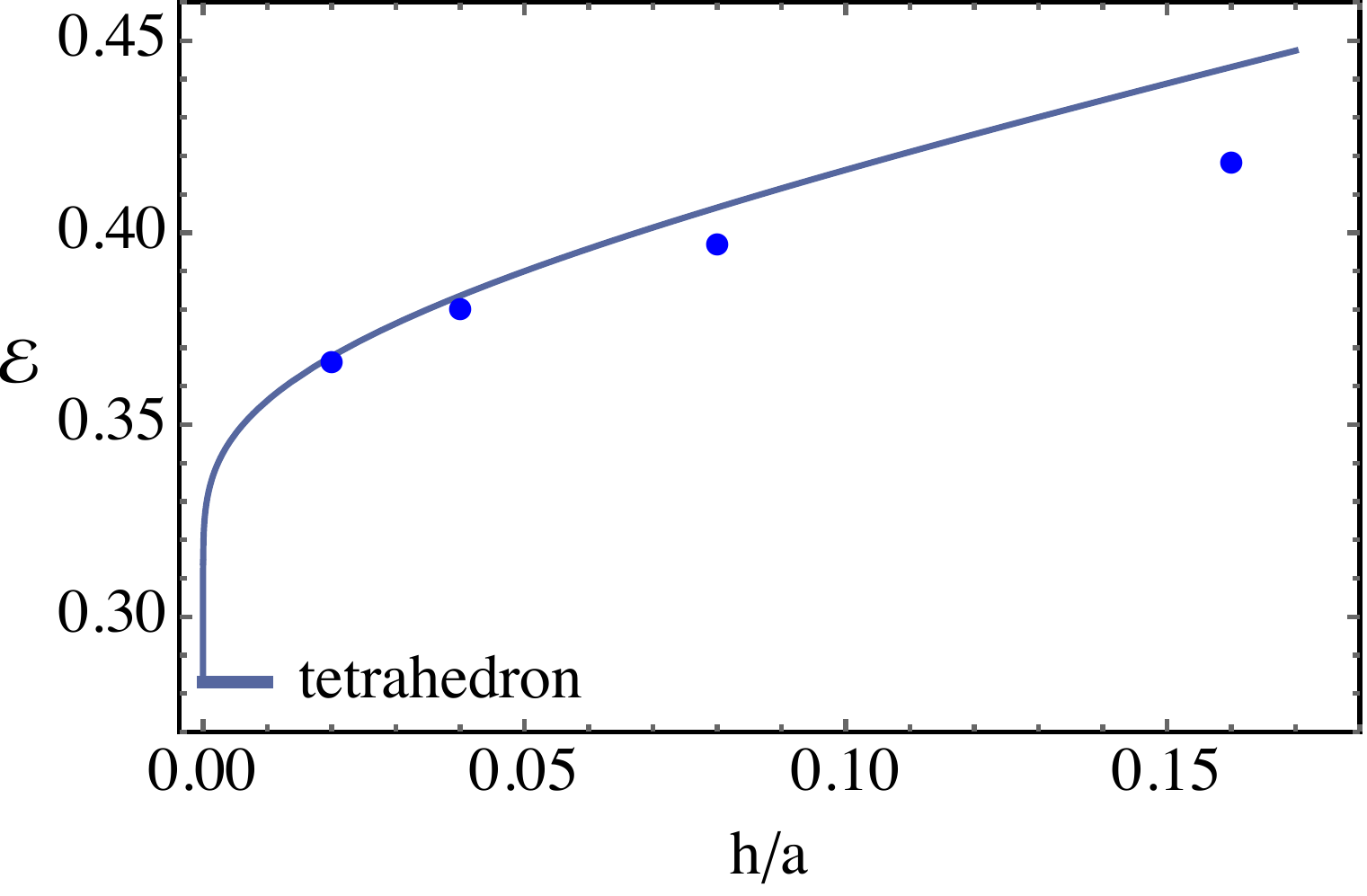}
\quad
\includegraphics[width=.45\textwidth,height=!]{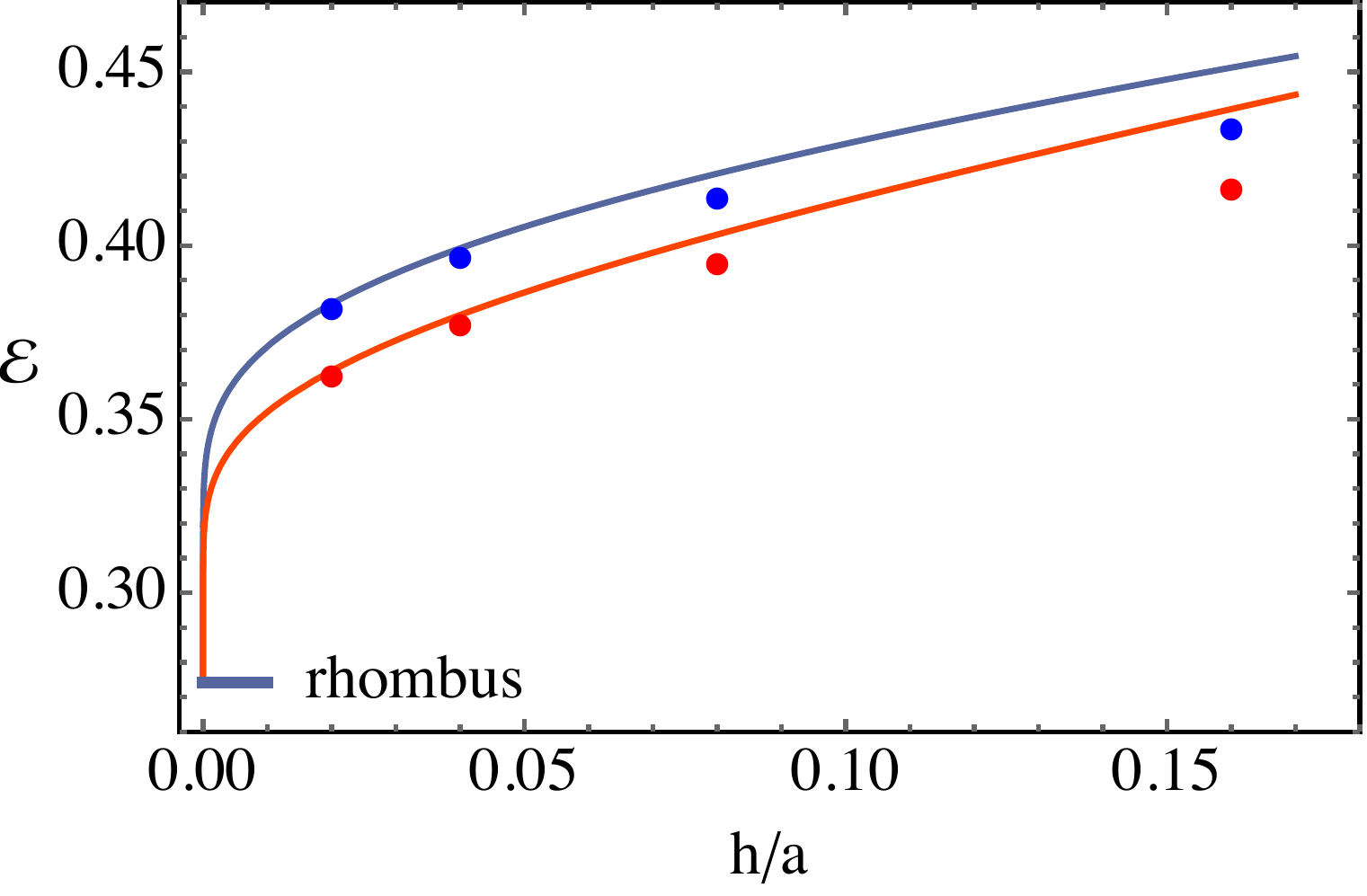}
\caption{
Left: energy of a tetrahedron cluster with one charge placed on one of the spheres.
Right: for a rhombus cluster with the charge placed on the pointed (blue) and blunt (red) corners. 
}
\label{fig:ETetra}
\end{figure}

Examples of energies $\calE$ found using $\bbH$ from the previous section are shown in Fig.~(\ref{fig:ETetra}), for a tetrahedron and a rhombus.
For the rhombus case, two sets of comparisons were made,
one having a charge placed on the sphere at pointed position and one at the blunt position.
The discrepancy for $h/a \geq 0.05$ is apparent and can be attributed to the weak $h$-dependence of $\bbH$.

\begin{figure}[bt]
\centering
\includegraphics[width=.45\textwidth,height=!]{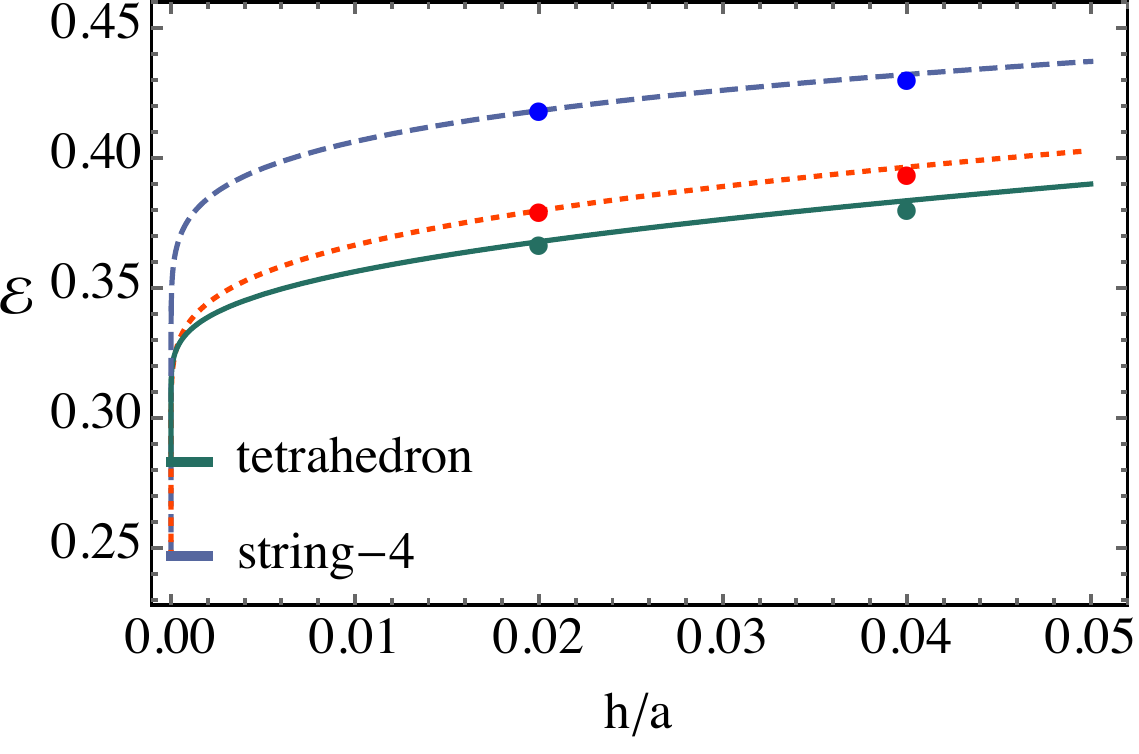}
\includegraphics[width=.45\textwidth,height=!]{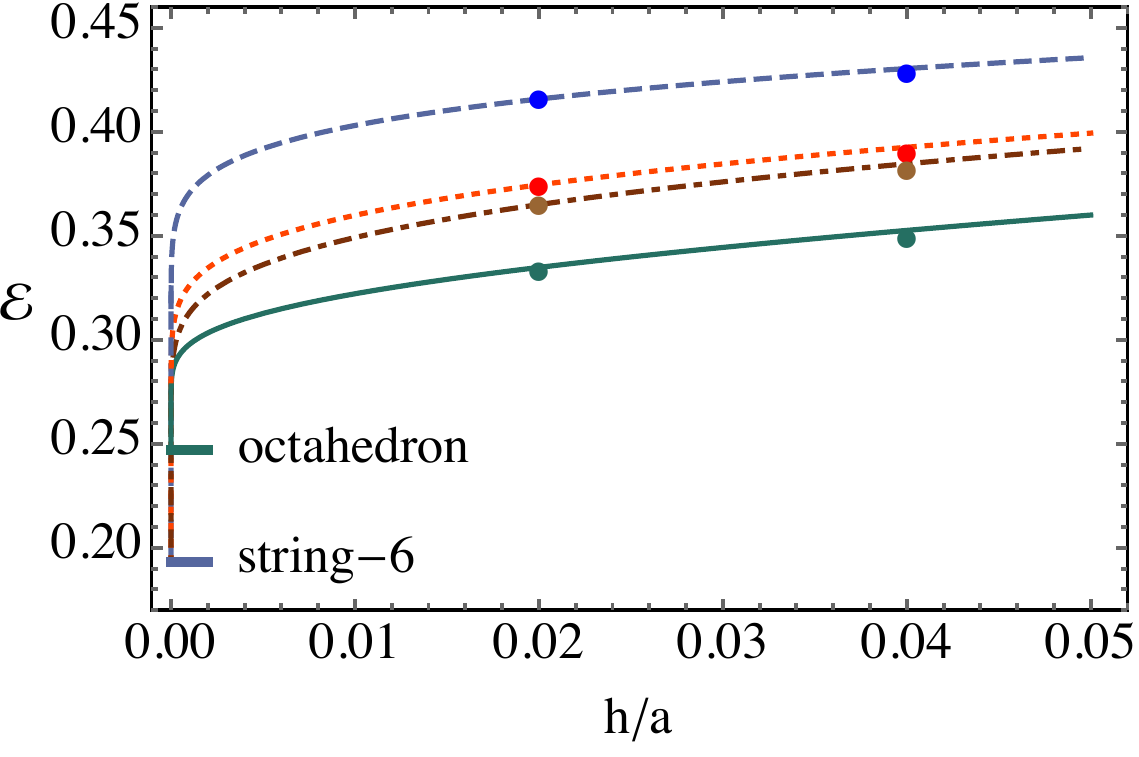}
\caption{
Energies of tetrahedron and octahedron clusters and their linear string correspondence,
with one charged sphere.
Left: tetrahedron (solid) and string (dashed).
upper curve: charge is on the sphere at the end of string;
middle curve: charge is on the second sphere.
Right: octahedron (solid) and string (dashed).
top curve: end sphere is charged;
second and third curves from top: second and third from end sphere is charged.
}
\label{fig:Estring}
\end{figure}

The results in Fig.~\ref{fig:Esize} show that the binding energy, attained at $h=0$,
for a cluster with extended structure is always lower than that of compact ones,
which suggests that a typical cluster configuration is always a linear string.
However, the $h$ dependences of compact and extended configurations are different.
Since contacts can only lower the energy,
the contact-energy correction tends to favor compact clusters with the most contacts.
Further, for large clusters with a single charged sphere (Appendix~\ref{sec:shape_appendix}) the contact energy leads to a lower net energy for compact clusters versus extended clusters. 
Fig.~\ref{fig:Estring} compares the energies between compact and extended configurations
for cluster of $4$ and $6$ spheres.
In both cases the compact configuration is always energetically favorable for the visible range of $h$.

Since the relevant Laplacian matrix $\tilde \bbL$ of a connected cluster is positive definite \cite{positiveNote}, 
moving spheres away from each other always raises the energy,
resulting in a logarithmically attractive potential well.
Consequently, we expect any types of clusters to be stable at sufficiently small separation.

\section{Discussion}
\label{sec:discussion}

The preceding sections have explored a peculiar type of Coulomb interaction arising from the charging constraints encountered at the spatial scales of nanoparticles.  Below we note the limitations of our work and suggest experimental situations where the interaction discussed here might nevertheless be relevant.  

In order to demonstrate the specific features our mechanism, we have considered the simplest example that shows the necessary features.  First, a cluster of spheres like those considered here has charge polarization extending beyond the induced dipoles normally considered.  Second, any excess charge on the cluster is dominated by single electron charges residing on one or another of the spheres.  Given these two features, one should observe the singular dependence on separation $h$ found above. Our main aim has been to show the form of this singularity and how its dependence on separation may be understood.

The relative electrostatic energies of different clusters are important for determining their relative abundance and stability.  For real experimental situations  the relative abundance of actual nanocluster shapes doubtless depends strongly on several other factors as well.  In real clusters, it is likely too simplistic to assume a single charge on a particular sphere; a number of charge distributions likely have significant probability.  If a single charge is present, it may reside on any sphere of a cluster that doesn't require an extra energy much higher than $k_B T$.  Thus in practice one may need to consider an average over several charge positions in order to determine the stability of a given cluster shape.

Our calculations have concentrated on the effects of the logarithmic singularity, important when the separation $h$ is much smaller than the sphere radii.  In cases where more accuracy is desired for larger separations, our scheme can be naturally extended by replacing the regular part $\bbH$ by a Taylor series $\bbH_0 + h \bbH_1 + \cdots$. 

One might expect that this attractive mechanism should extend beyond conducting spheres to dielectric spheres, especially if the dielectric contrast is large.  However, the concentration of charge near a contact is qualitatively weaker for dielectrics than for conductors.  For dielectrics in an external field, the charge density remains finite at contact; it does not diverge as in the conducting case.

Though we have only treated the specific case of clusters of spheres of equal size, the effects explored here apply generally to convex conductors.  When any two smoothly curving conductors approach each other, the Derjaguin argument of Sec.~\ref{sec:twospheres} implies a logarithmically diverging mutual capacitance, whose $c(h)$ depends only on the mean curvatures of the two adjacent surfaces.  

In real materials a net charge on a cluster is only created in combination with a countercharge elsewhere.  In practice these countercharges may lie close to  the cluster and thus modify the coulomb energy significantly.  Thus our results only apply when the screening length due to external charge is larger than the cluster.  

Naturally real clusters like those of Fig.~\ref{fig:micrographs} experience other forms of interaction unrelated to net charge on the cluster.  The organic coronas \cite{TalapinNature2006} used to to stabilize the particles exert interparticle forces.  So do steric interactions with other neighboring nanoparticles.  Dispersion forces and solvent-specific chemical interactions are also present.  In order to make reliable predictions of cluster shapes, one would need to add these conventional interactions to the charge-induced interactions considered here.  

Experimental consequences of our clustering mechanism could potentially be found in the binary lattices like Fig.~\ref{fig:micrographs} that motivated our study.  If our mechanism is important, one expects (a) cluster shapes with lower electrostatic energy as calculated above should be relatively more prevalent, and (b) particles with a thicker ligand layer should be less strongly bound but have greater preference for specific charge sites.  Still, the number of competing effects that determine the specific cluster shapes precludes any decisive predictions. 

Other simpler systems give a brighter prospect for decisive predictions.
One such system is a dilute dispersion of nanoparticles in a nonpolar solvent \cite{Fernandez:2009fk}.  One may induce charge separation by adding large counterions to the dispersion \cite{Sainis:2008uq}.
Then any nanoparticle with a net charge will attract neutral nanoparticles via the mechanism described above.  If the counterions are sufficiently large and distant, their effects can be made minor.  Then one expects to observe clusters with relative abundance dictated in thermal equilibrium by the electrostatic binding energies described above.  

\section{Conclusion}
\label{sec:conclusion}  

We have shown that the electrostatic energy of a cluster of spherical conductors has a novel form when one conductor is charged and their separations are small.  In the limit of small separations the energy is finite, but the corrections to this limit are logarithmically singular.  Thus for real clusters where the separation is nonzero, it is important to know the singular contribution.  Both the limiting energy and the corrections can be expressed in terms of non-singular operations. It appears from our numerical examples that these small separations can have a significant impact on the binding of the clusters. In certain situations as noted above, this distinctive form of binding could be significant in determining the prevalent cluster shapes. 

\begin{acknowledgments}
The authors are grateful to Prof. Dmitri Talapin for numerous discussions of his experimental findings on nanoparticle self-assembly.  We thank Toan Nguyen, Eric Dufresne and Naomi Oppenheimer for insightful discussions.   We thank Alexander Moore, author of Ref.~\citenum{AMooreArXiV2010}, for extensive discussions and the use of his algebraic code. We thank Jason Merrill for pointing out a serious error in a previous draft.  We thank the Aspen Center for Physics for hospitality during part of this work.  
This work was supported by the National Science Foundation's MRSEC
Program under Award Numbers DMR-0820054 and DMR-142070.
JQ was supported by the Kadanoff-Rice fellowship through the University of Chicago.
\end{acknowledgments}

\appendix
\section{Shape-dependence of cluster energy}
\label{sec:shape_appendix}

In this appendix we estimate how the Coulomb energy of a cluster depends on its overall shape: compact vs extended.  We consider a large cluster of $n$ spheres under two extremes of compactness. On the one hand we consider the cluster of least compactness, where all the spheres are extended along a one-dimensional line.  On the other hand, we consider the state of maximal compactness in which the spheres form a spherical aggregate of maximum density.   As noted in the main text, the equipotential part of the energy favors extended structures. Here we focus on the leading logarithmic correction to the binding energy $\calE(h)$ from Eq.~(\ref{eq:oldcalE}). We denote it as $\calE'$,
$$
\calE' = \frac 1 {c(h)}~ \vQ' \cdot \tilde\bbL^{-1} \vQ' .
$$
Our interest is in the case where the charge $\vQ$ is concentrated in a single sphere.  

In both of these clusters, one may use a continuum approach to characterize $\bbL$.  The $\bbL$ has a simple interpretation in terms of a quantum system.  In this system one replaces each sphere by a site and each contact by a connecting junction.  The $\bbL$ matrix is then the Hamiltonian of a quantum particle in this system and its eigenstates are the energy levels.  For a homogeneous solid these eigenstates are the well-known tight-binding states of solid state physics \cite{Kittel:2005rz}.  The $n$ eigenstates $\vk$ are normalized plane waves of wavevector ${\underline k}$ and eigenvalues of order $k^2$.  Using this fact we may write $\calE'$ as 
$$
\calE' \goesas \frac 1 {c(h)}~ \sum_{\underline k} \vQ' \cdot \vk \quad \frac 1 {k^2}\quad \vk \cdot \vQ' ~.
$$
Here sum goes over the $n$ distinct wave states compatible with the boundary conditions.  The vectors $\vQ'$ are constructed to have vanishing projection on the $k=0$ state, so $k=0$ is omitted from this sum.  

To compute the sum, we need to know the dot products $\vQ'\cdot \vk$.  The $\vQ'$ is the sum of two parts: $\vQ$ and $\vQ_e$. We first consider the $\vQ$ part, which vanishes except on a particular sphere. It is the discrete analog of a delta function in space.  Accordingly it has an equal dot product onto all the $k$ eigenstates, each of order $n^{-1/2}$.  Thus we may treat these dot products as constants in the sum.  We may also replace the $\sum_k$ by the integral $L^d\integral d^dk$ for a d-dimensional cluster of linear size $L$.  Then $\calE'$ simplifies to 
$$
\calE' \goesas \frac 1 {c(h)}~ (\vk \cdot \vQ)^2 ~L^d ~\left ( \integral_{k_{min}}^{k_{max}} ~\frac 1 {k^2}~k^{d-1}dk \right )  ~.
$$
Here $k_{min}\goesas L^{-1}$ and $k_{max}\goesas L^0$.  

For a one-dimensional cluster the integral is dominated by the lower limit and 
$$
\calE' \goesas \frac 1 {c(h)}~ (\vk \cdot \vQ)^2 ~L ~ (L) ~.   
$$
Since $(\vk \cdot \vQ)^2 \goesas 1/n$ and $L \goesas n$, we have $\calE' \goesas n$.  

For a three-dimensional cluster the integral is dominated by the upper limit and 
$$
\calE' \goesas \frac 1 {c(h)}~ (\vk \cdot \vQ)^2 ~L^3   ~.
$$
Using $(\vk \cdot \vQ)^2 \goesas 1/n$ and $L \goesas n^{1/3}$, we conclude $\calE' \goesas n^0$.  

We now consider the effect of the $\vQ_e$ part of $\vQ'$.  For both clusters $\vQ_e$ is concentrated at the outer boundary.  It thus has significant Fourier components at large $k$.  However, this charge concentration is in any case qualitatively weaker than the complete concentration found in $\vQ$.  Accordingly we expect the $\vQ_e$ part of $\vQ'$ to have a minor effect and the scaling estimates for $\calE'$ to hold for the full $\vQ'$ as for the $\vQ$.  

The foregoing estimates indicate a qualitative difference in $\calE'$ in the two cases.  This positive energy diverges with $n$ for the extended cluster but remains finite for the compact cluster.  It disfavors the extended cluster.  This contrasts with the equipotential part of $\calE$, which favors extended clusters.  This equipotential energy $\calE_e$ is of order $\log n/n$ for extended clusters and of order $1/L \goesas n^{-1/3} $ for compact clusters.  The total energy $\calE = \calE_e + \calE'$ thus favors compact clusters, in contrast to the ``leading" $\calE_e$ alone.


\begin{thebibliography}{19}
\bibitem [{pos()}]{positiveNote}%
  \BibitemOpen
  \href@noop {} {}\bibinfo {note} {The ``reduced energy'' $\vV \cdot \bbL \vV$
  is the sum of positive contact energies. Since the cluster is connected, if
  any two potentials $V_i, V_j$ are unequal there is some contact linking
  sphere $i$ to sphere $j$ whose contact voltage is nonzero. and this gives a
  positive contribution to $\vV \cdot \bbL \vV$.}\BibitemShut {Stop}%
\bibitem [{inv()}]{invertibleNote}%
  \BibitemOpen
  \href@noop {} {}\bibinfo {note} {If the equation is not invertible it must
  have some null vector. That is there must be some $\vV'$ for which $\bbH \vV'
  = -c(h) \tilde\bbL \vV'$. Since the dimensions of $\bbH$ and $\tilde\bbL$ are
  finite, this is possible only for a finite set of values of $c(h)$. If $c(h)$
  is larger than the largest of these, there is no null $\vV'$ and the equation
  is invertible.}\BibitemShut {Stop}%
\end{thebibliography}
\end{document}